\def\dps{\displaystyle}
\def\be{\begin{equation}}
\def\ee{\end{equation}}
\def\ba{\begin{array}}
\def\ea{\end{array}}

\documentclass[aps,amsmath,amssymb,amsfonts,showpacs,twocolumn,pra]{revtex4}
\usepackage{graphicx}
\usepackage{caption2}
\def\qed{\leavevmode\unskip\penalty9999 \hbox{}\nobreak\hfill
     \quad\hbox{\leavevmode  \hbox to.77778em{%
               \hfil\vrule   \vbox to.675em%
               {\hrule width.6em\vfil\hrule}\vrule\hfil}}
     \par\vskip3pt}

\newtheorem{theorem}{Theorem}
\newtheorem{corollary}{Corollary}

\begin{document}
\title{Uncertainty relations based on mutually unbiased measurements}
\author{Bin Chen$^{1}$}
\author{Shao-Ming Fei$^{1,2}$}

\affiliation{$^1$School of Mathematical Sciences, Capital Normal University,
Beijing 100048, China\\
$^2$Max-Planck-Institute for Mathematics in the Sciences, 04103
Leipzig, Germany}

\begin{abstract}
We derive uncertainty relation inequalities according to the mutually unbiased measurements.
Based on the calculation of the index of coincidence of probability distribution given by $d+1$ MUMs on any density operator $\rho$ in $\mathbb{C}^{d}$,
both state-dependent and state-independent forms of lower entropic bounds are given.
Furthermore, we formulate uncertainty relations for MUMs in terms of R\'{e}nyi and Tsallis entropies.
\end{abstract}

\maketitle

\section{Introduction}
Uncertainty relation and complementarity principle are two key concepts in both quantum mechanics and quantum information theory. The best known form of Heisenberg's uncertainty relations, given by Robertson \cite{Rob}, states that if one prepares a large number of copies of a state $|\psi\rangle$, and measures two observables $O_{1}$ and $O_{2}$ individually, then the standard deviation $\Delta(O_{i})$ of $O_{i}$, defined as $\Delta(O_{i})=\sqrt{\langle O_{i}^{2}\rangle-\langle O_{i}\rangle^{2}}, i=1,2$, satisfy the following inequality
$$
\Delta O_{1}\Delta O_{2}\geq\frac{1}{2}|\langle\psi|[O_{1},O_{2}]|\psi\rangle|,
$$
where $[O_{1},O_{2}]$ is the commutator of $O_{1}$ and $O_{2}$. As a consequence of uncertainty relations, the complementarity principle claims that it is impossible to simultaneously determine the exact values of the two non-commuting observables.

Uncertainty relations can be also characterized in terms of entropies. The so-called entropic uncertainty relations were originally pointed out by Deutsch \cite{Deu} and later improved by Maassen and Uffink \cite{Maa}, who derived an entropic uncertainty relation for a pair of mutually unbiased bases (MUBs). Two orthonormal bases in $d$-dimensional complex vector space $\mathbb{C}^{d}$ are said to be mutually unbiased if the absolute values of the inner products of any basic vector in one basis and any basic vector in another basis are $1/\sqrt{d}$. A set of orthonormal bases is called a set of mutually unbiased bases if every pair of bases in the set are mutually unbiased. The maximum number $N(d)$ of MUBs in a set of mutually unbiased bases is no more than $d+1$, and $N(d)=d+1$ when $d$ is a prime power \cite{Woo}. But when $d$ is a composite number, $N(d)$ remains unknown \cite{Durt}. MUBs play an important role in the investigation of uncertainty relations \cite{Ivo, Sa, SaRu, Wu, Weh}. In Ref.\cite{Wu}, assuming the existence of $M$ MUBs, the authors presented a number of inequalities which lead to tighter and more general entropic uncertainty relations than the previous ones. Recently, Kalev and Gour generalize the concept of MUBs to mutually unbiased measurements (MUMs) \cite{Kal}. They show that there exists a complete set of $d+1$ MUMs for arbitrary $d$, which can be explicitly constructed. MUMs can also be used to derive entropic uncertainty relations, and a state-independent formulation is obtained in \cite{Kal}.

Similar to mutually unbiased bases, another important concept in quantum information theory is the symmetric informationally complete positive operator-valued measurements (SIC-POVMs). A set of $d^{2}$ operators in $\mathbb{C}^{d}$ is said to be a SIC-POVM, if it is a POVM in which all elements are of the form $d^{-1}$ times a rank-one projector, and the operator inner products of any two elements are the same. Although in a number of low-dimensional cases, the existence of SIC-POVMs has been proved analytically, and numerically for all dimensions up to 67, it is still unknown whether or not SIC-POVMs exist for arbitrary $d$ \cite{Sco}. In Ref. \cite{Kal2}, Kalev and Gour generalize the concept of SIC-POVMs to general symmetric informationally complete (SIC) measurements. They construct the set of all general SIC measurements, in which the elements need not be of rank-one. Like MUBs and MUMs, SIC-POVMs and general SIC measurements are also useful in studying uncertainty relations \cite{RastEur, Rastnotes}.

Besides Shannon entropy, other entropies also play key roles in classical and quantum information theory, especially in the investigation of entropic uncertainty relations. In Ref. \cite{RastEur}, the author formulated uncertainty relations for MUBs and SIC-POVMs in terms of R\'{e}nyi and Tsallis entropies. Lower entropic bounds for general SIC measurements in terms of such entropies are derived in Ref. \cite{Rastnotes}.

In this paper, we first calculate the so-called index of coincidence of probability distribution given by a complete set of mutually unbiased measurements on any density operator $\rho$. This general result, including two special cases in Ref. \cite{Wu,Kal}, can be used to derive a state-dependent entropic uncertainty relation (see theorem 2). The previous state-independent entropic uncertainty inequality obtained in Ref. \cite{Kal} can be deduced from our result, accounting to the fact that $\mathrm{Tr}(\rho^{2})\leq1$. Furthermore, we provide some state-dependent and state-independent uncertainty relations based on MUMs, as well as a single general SIC measurements by using the Harremo\"{e}s-Tops{\o}e theorem, an approach used in Ref. \cite{Wu} in deriving entropic uncertainty relations for $M$ MUBs in $\mathbb{C}^{d}$. At last, we discuss uncertainty relations in terms of R\'{e}nyi and Tsallis entropies for MUMs, as for general SIC measurements in Ref. \cite{Rastnotes}.

\section{index of coincidence for MUMs and entropic uncertainty relations}
Let us first recall some basic notions of mutually unbiased measurements \cite{Kal}. Two POVM  measurements on $\mathbb{C}^{d}$, $\mathcal{P}^{(b)}=\{P_{n}^{(b)}\}_{n=1}^{d}$, $b=1,2$, are said to be mutually unbiased measurements if
\begin{equation}
\begin{split}
\mathrm{Tr}(P_{n}^{(b)})&=1,\\
\mathrm{Tr}(P_{n}^{(b)}P_{n'}^{(b')})&=\frac{1}{d},~~~b\neq b',\\
\mathrm{Tr}(P_{n}^{(b)}P_{n'}^{(b)})&=\delta_{n,n'}\,\kappa+(1-\delta_{n,n'})\frac{1-\kappa}{d-1},
\end{split}
\end{equation}
where $\frac{1}{d}<\kappa\leq1$, and $\kappa=1$ if and only if all $P_{n}^{(b)}$'s are rank one, which gives rise to a complete set of $d+1$ mutually unbiased bases. Unlike the existence of a complete set of MUBs, such MUMs do exist for arbitrary $d$, and can be explicitly constructed \cite{Kal}. Let $\{F_{n,b}:n=1,2,\ldots,d-1,b=1,2,\ldots,d+1\}$ be a set of $d^{2}-1$ Hermitian, traceless operators acting on $\mathbb{C}^{d}$, satisfying $\mathrm{Tr}(F_{n,b}F_{n',b'})=\delta_{n,n'}\delta_{b,b'}$. Define $d(d+1)$ operators
\begin{equation}\label{2}
F_{n}^{(b)}=
\begin{cases}
   F^{(b)}-(d+\sqrt{d})F_{n,b},&n=1,2,\ldots,d-1;\\[2mm]
   (1+\sqrt{d})F^{(b)},&n=d,
\end{cases}
\end{equation}
where $F^{(b)}=\sum_{n=1}^{d-1}F_{n,b}$, $b=1,2,\ldots,d+1.$ Then the operators
\begin{equation}\label{3}
P_{n}^{(b)}=\frac{1}{d}I+tF_{n}^{(b)},
\end{equation}
with $b=1,2,\cdots,d+1,n=1,2,\cdots,d,$ form $d+1$ MUMs, as long as $t$ is chosen such that $P_{n}^{(b)}\geq0$. Moreover, any $d+1$ MUMs can be expressed in such form.

Note that the operators $F_{n}^{(b)}$ satisfy the following properties
\begin{equation}\label{4}
\begin{split}
\mathrm{Tr}(F_{n}^{(b)}F_{n'}^{(b)})&=(1+\sqrt{d})^{2}[\delta_{nn'}(d-1)-(1-\delta_{nn'})],\\
\sum_{n=1}^{d}F_{n}^{(b)}&=0,\\
\mathrm{Tr}(F_{n}^{(b)}F_{n'}^{(b')})&=0,~~~\forall b\neq b',~ \forall n,n'=1,2,\ldots,d.
\end{split}
\end{equation}
The parameter $\kappa$ is given by
\begin{equation}\label{5}
\kappa=\frac{1}{d}+t^{2}(1+\sqrt{d})^{2}(d-1).
\end{equation}

To derive entropic uncertainty relations for MUMs, we have to calculate the so-called index of coincidence of probability distribution given by $d+1$ MUMs on any density operator $\rho$ acting on $\mathbb{C}^{d}$. For a given probability distribution $\textbf{p}=(p_{1},p_{2},\ldots p_{d})$, the index of coincidence is defined by $C(\textbf{p})=\sum_{i=1}^{d}p_{i}^{2}$ \cite{HT}. Let $\{\mathcal{P}^{(b)}\}_{b=1}^{d+1}$ be a set of $d+1$ MUMs on $\mathbb{C}^{d}$ with the parameter $\kappa$, where $\mathcal{P}^{(b)}=\{P_{n}^{(b)}\}_{n=1}^{d},b=1,2,\ldots,d+1$. Let $p_{n}^{(b)}$ denotes the probability of the outcome when measuring $\rho$ with $P_{n}^{(b)}$, i.e. $p_{n}^{(b)}=\mathrm{Tr}(P_{n}^{(b)}\rho)$.

\begin{theorem} Denote $C(\kappa,\rho)$ the index of coincidence for probability distribution $\{p_{n}^{(b)}\}$. We have
\begin{equation}\label{t1}
C(\kappa,\rho)=\frac{(d\kappa-1)[d\mathrm{Tr}(\rho^{2})-1]+d^{2}-1}{d(d-1)}.
\end{equation}
\end{theorem}

\emph{Proof.} Any quantum state can be written as \cite{Kal},
$$
\rho=\frac{1}{d}I+\sum_{b=1}^{d+1}\sum_{n=1}^{d}r_{n}^{(b)}F_{n}^{(b)}.
$$
Using formulae (\ref{4}), one can easily get
$$\mathrm{Tr}(\rho^{2})=\frac{1}{d}+(1+\sqrt{d})^{2}\sum_{b=1}^{d+1}[d\sum_{n=1}^{d}(r_{n}^{(b)})^{2}-(\sum_{n=1}^{d}r_{n}^{(b)})^{2}].$$
From the construction of MUMs (\ref{3}), we have
$$
p_{n}^{(b)}=\mathrm{Tr}(P_{n}^{(b)}\rho)=\frac{1}{d}+t(1+\sqrt{d})^{2}(dr_{n}^{(b)}-\sum_{n'=1}^{d}r_{n'}^{(b)}).
$$
Therefore
\begin{eqnarray*}
C(\kappa,\rho) & = & \sum_{b=1}^{d+1}\sum_{n=1}^{d}(p_{n}^{(b)})^{2}\\
& = & \frac{d+1}{d}+t^{2}(1+\sqrt{d})^{4}\sum_{b=1}^{d+1}\sum_{n=1}^{d}(d\,r_{n}^{(b)}-\sum_{n'=1}^{d}r_{n'}^{(b)})^{2}\\
& & +2t(1+\sqrt{d})^{2}\sum_{b=1}^{d+1}\sum_{n=1}^{d}(r_{n}^{(b)}-\frac{1}{d}\sum_{n'=1}^{d}r_{n'}^{(b)})\\
& = & \frac{d+1}{d}+t^{2}(1+\sqrt{d})^{4}\sum_{b=1}^{d+1}\sum_{n=1}^{d}\left[d^{2}\,(r_{n}^{(b)})^{2}\right.\\
& & \left.+(\sum_{n'=1}^{d}r_{n'}^{(b)})^{2}-2\,d\,r_{n}^{(b)}\sum_{n'=1}^{d}r_{n'}^{(b)}\right]\\
& = & \frac{d+1}{d}+t^{2}(1+\sqrt{d})^{4}\sum_{b=1}^{d+1}\left[d^{2}\sum_{n=1}^{d}(r_{n}^{(b)})^{2}\right.\\
& & \left.-d(\sum_{n'=1}^{d}r_{n'}^{(b)})^{2}\right]\\
& = & \frac{d+1}{d}+t^{2}(1+\sqrt{d})^{2}(d\,\mathrm{Tr}(\rho^{2})-1).
\end{eqnarray*}
From (\ref{5}) we get (\ref{t1}). \quad $\Box$

If $\kappa=1$, the set of $d+1$ MUMs $\{\mathcal{P}^{(b)}\}_{b=1}^{d+1}$ are reduced to a complete set of MUBs, and $C(1,\rho)=\mathrm{Tr}(\rho^{2})+1$, which gives rise to the result in Ref. \cite{Wu}. If $\rho$ is a pure state, then $C(\kappa,\rho)=\kappa+1$, which gives rise to the result in Ref. \cite{Kal}.

Now we can derive uncertainty relations by using the theorem. We first consider the Shannon entropy defined by $H(\textbf{p})=-\sum_{j=1}^{d}p_{j}\log_{2}p_{j}$, where the probability distribution $\textbf{p}=(p_{1},p_{2},\ldots p_{d})$. Entropic uncertainty relations in terms of the R\'{e}nyi and Tsallis entropies will be discussed in the next section.

\begin{theorem}
For a set of $d+1$ MUMs $\{\mathcal{P}^{(b)}\}_{b=1}^{d+1}$ on $\mathbb{C}^{d}$ with the parameter $\kappa$, we have the following state-dependent entropic uncertainty relation:
$$
\frac{1}{d+1}\sum_{b=1}^{d+1}H(\mathcal{P}^{(b)}|\rho)\geq\log_{2}\frac{d+1}{C(\kappa,\rho)},
$$
where $H(\mathcal{P}^{(b)}|\rho)$ denotes the Shannon entropy of the probability distribution generated by $\mathcal{P}^{(b)}$ with respect to $\rho$, and $C(\kappa,\rho)$ is given by (\ref{t1}).
\end{theorem}

\emph{Proof.} As $\mathcal{P}^{(b)}=\{P_{n}^{(b)}\}_{n=1}^{d},b=1,2,\ldots,d+1$, and $p_{n}^{(b)}=\mathrm{Tr}(P_{n}^{(b)}\rho)$. From the concavity of the log function \cite{Kal}, we have the following inequality,
\begin{eqnarray*}
\frac{1}{d+1}\sum_{b=1}^{d+1}H(\mathcal{P}^{(b)}|\rho) & \geq &
-\frac{1}{d+1}\sum_{b=1}^{d+1}\log_{2}\sum_{n=1}^{d}(p_{n}^{(b)})^{2}\\
& \geq & -\log_{2}\left[\frac{1}{d+1}\sum_{b=1}^{d+1}\sum_{n=1}^{d}(p_{n}^{(b)})^{2}\right]\\
& = & \log_{2}\frac{d+1}{C(\kappa,\rho)},
\end{eqnarray*}
where $C(\kappa,\rho)$ is given by theorem 1. \quad $\Box$

Accounting to the fact that $\mathrm{Tr}(\rho^{2})\leq1$, we can derive the state-independent entropic uncertainty relation,
\begin{equation}\label{6}
\frac{1}{d+1}\sum_{b=1}^{d+1}H(\mathcal{P}^{(b)}|\rho)\geq\log_{2}\frac{d+1}{\kappa+1},
\end{equation}
which was derived in Ref. \cite{Kal}. If $\kappa=1$, our theorem coincides with the result in Ref. \cite{Wu}.

In Ref. \cite{Wu}, assuming that there exist $M$ MUBs in $\mathbb{C}^{d}$, the authors derived some entropic uncertainty relations by using Harremo\"{e}s-Tops{\o}e theorem \cite{HT}. This method is also valid for MUMs. For a given probability distribution $\textbf{p}=(p_{1},p_{2},\ldots p_{d})$, the Harremo\"{e}s-Tops{\o}e theorem tells us that the Shannon entropy $H(\textbf{p})$ and the index of coincidence $C(\textbf{p})=\sum_{i=1}^{d}p_{i}^{2}$ satisfy the following inequality for any integer $1\leq x\leq d-1$:
\begin{eqnarray*}
H(\textbf{p})&\geq&[(x+1)\log_{2}(x+1)-x\log_{2}x]\\[1mm]
& & -C(\textbf{p})x(x+1)[\log_{2}(x+1)-\log_{2}x].
\end{eqnarray*}
Following the above notation, we have
\begin{eqnarray*}
\sum_{b=1}^{d+1}H(\mathcal{P}^{(b)}|\rho)&\geq&(d+1)[(x+1)\log_{2}(x+1)-x\log_{2}x]\\
& & -C(\kappa,\rho)x(x+1)[\log_{2}(x+1)-\log_{2}x].
\end{eqnarray*}
Let $C$ be an upper bound for $C(\kappa,\rho)$. Then
\begin{eqnarray*}
\sum_{b=1}^{d+1}H(\mathcal{P}^{(b)}|\rho)&\geq&(d+1)[(x+1)\log_{2}(x+1)-x\log_{2}x]\\
& & -C x(x+1)[\log_{2}(x+1)-\log_{2}x]\\[1mm]
&=& (d+1-C x)(x+1)\log_{2}(x+1)\\[1mm]
& & -[d+1-C(x+1)]x\log_{2}x\\
&:=& f(x).
\end{eqnarray*}
It has been proved that \cite{Wu} $f(x)$ gets its maximal value at $x=\lfloor\frac{d+1}{C}\rfloor$. Therefore, we have the following entropic uncertainty inequality:
\begin{theorem}
$$
\sum_{b=1}^{d+1}H(\mathcal{P}^{(b)}|\rho)\geq a\,C\,(h+1)\log_{2}(h+1)+(1-a)\,C\,h\log_{2}h,
$$
where $C$ is an upper bound for $C(\kappa,\rho)$, $h=\lfloor\frac{d+1}{C}\rfloor$ and $a=\frac{d+1}{C}-h$.
\end{theorem}

We can choose $C=1+\kappa$ since $\mathrm{Tr}(\rho^{2})\leq1$. Then we obtain the following state-independent inequality which is stronger (as noted in \cite{Wu}) than (\ref{6}).
\begin{corollary}
$$
\ba{l}
\dps\frac{1}{d+1}\sum_{b=1}^{d+1}H(\mathcal{P}^{(b)}|\rho)\geq\log_{2}h\\
\dps~~~~~~~~~~~~~~~~~~+\left[1-(\frac{\kappa+1}{d+1})h\right](h+1)\log_{2}(1+\frac{1}{h}),
\ea
$$
where $h=\lfloor\frac{d+1}{\kappa+1}\rfloor.$
\end{corollary}

We now briefly discuss uncertainty relations for a single general SIC measurements by using the Harremo\"{e}s-Tops{\o}e theorem. A set of $d^{2}$ positive-semidefinite operators $\mathcal{P}=\{P_{j}\}_{j=1}^{d^{2}}$ on $\mathbb{C}^{d}$ is said to be a general SIC measurements, if \\
$$
\ba{l}
(1)~ \dps\sum_{j=1}^{d^{2}}P_{j}=I,\\[5mm]
(2)~ \dps\mathrm{Tr}(P_{j}^{2})=a,\\[2mm]
~~~~~\dps\mathrm{Tr}(P_{j}P_{k})=\frac{1-da}{d(d^{2}-1)},~\forall j,k\in\{1,2,\ldots,d^{2}\},~j\neq k,
\ea
$$
where $I$ is the identity operator, the parameter $a$ satisfies $\frac{1}{d^{3}}<a\leq\frac{1}{d^{2}}$, and
$a={1}/{d^{2}}$ if and only if all $P_{j}$ are rank one, which gives rise to a SIC-POVM.
It can be shown that $\mathrm{Tr}(P_{j})=\frac{1}{d}$ for all $j$ \cite{Kal2}.

Let $\rho$ be an density operator in $\mathbb{C}^{d}$ and $p_{j}=\mathrm{Tr}(P_{j}\rho)$. The index of coincidence for general SIC measurements has been calculated as \cite{Rastnotes}
$$
C(a,\rho)=\sum_{j=1}^{d^{2}}p_{j}^{2}=\frac{(ad^{3}-1)\mathrm{Tr}(\rho^{2})+d(1-ad)}{d(d^{2}-1)}.
$$
Let $C$ be an upper bound of $C(a,\rho)$, and $x$ be any integer such that $1\leq x\leq d^{2}-1$. By using the Harremo\"{e}s-Tops{\o}e theorem, we get
\begin{eqnarray*}
H(\mathcal{P}|\rho)&\geq&[(x+1)\log_{2}(x+1)-x\log_{2}x]\\
& &-Cx(x+1)[\log_{2}(x+1)-\log_{2}x].
\end{eqnarray*}
The right-hand side of the above inequality reaches its maximal value at $x=\lfloor\frac{1}{C}\rfloor$ \cite{Wu}. Thus we have the following state-dependent uncertainty relation for a single general SIC measurements:
$$
\sum_{b=1}^{d+1}H(\mathcal{P}^{(b)}|\rho)\geq a\,C\,(h+1)\log_{2}(h+1)+(1-a)\,C\,h\log_{2}h,
$$
where $h=\lfloor\frac{1}{C}\rfloor$ and $a=\frac{1}{C}-h$. To obtain the state-independent form, we only need to set $C=\frac{ad^{2}+1}{d(d+1)}$ in the above inequality since $\mathrm{Tr}(\rho^{2})\leq1$.

\section{Uncertainty relations for MUMs in terms of R\'{e}nyi and Tsallis entropies}
In this section, we discuss some lower entropic bounds for mutually unbiased measurements in terms of R\'{e}nyi and Tsallis entropies.

For a given probability distribution $\textbf{p}=(p_{1},p_{2},\ldots p_{d})$, the R\'{e}nyi entropy is defined by \cite{Renyi}
$$
R_{\alpha}(\textbf{p})=\frac{1}{1-\alpha}\ln(\sum_{j=1}^{d}p_{j}^{\alpha}),
$$
where the parameter $\alpha>0$ and $\alpha\neq1$. When $\alpha\rightarrow1$, one gets the standard Shannon entropy $H(\textbf{p})=-\sum_{j=1}^{d}p_{j}\ln p_{j}$ (here we choose the natural logarithm $\ln$ in stead of $\log_{2}$). There are two more special cases which are respectively useful in studying uncertainty relations and cryptography: $\alpha=2$ gives rise to the so-called collision entropy \cite{Ball}
$$
R_{2}(\textbf{p})=-\ln(\sum_{j=1}^{d}p_{j}^{2}),
$$
and when $\alpha\rightarrow\infty$, one has the min-entropy \cite{Ng}
$$
R_{\infty}(\textbf{p})=-\ln(\max p_{j}).
$$

For $\alpha\in[2,\infty)$, it has been shown that \cite{RastEur}
$$
R_{\alpha}(\textbf{p})\geq\frac{\alpha}{2(1-\alpha)}\ln C(\textbf{p}),
$$
where $C(\textbf{p})$ is the index of coincidence of $\textbf{p}$. Let $\{\mathcal{P}^{(b)}\}_{b=1}^{d+1}$ be a set of $d+1$ MUMs on $\mathbb{C}^{d}$ with the parameter $\kappa$. Accounting to the convexity of the function $f(x)=(1-\alpha)^{-1}\ln x$ for $\alpha\geq2$ \cite{RastEur}, we obtain the following state-dependent uncertainty relation for MUMs in terms of R\'{e}nyi entropy:
$$
\frac{1}{d+1}\sum_{b=1}^{d+1}R_{\alpha}(\mathcal{P}^{(b)}|\rho)\geq
\frac{\alpha}{2(1-\alpha)}\ln\frac{C(\kappa,\rho)}{d+1},
$$
where $C(\kappa,\rho)$ is given by (\ref{t1}), $\alpha\geq2.$ Taking into account that $\mathrm{Tr}(\rho^{2})\leq1$,
we have the state-independent inequality,
$$
\frac{1}{d+1}\sum_{b=1}^{d+1}R_{\alpha}(\mathcal{P}^{(b)}|\rho)\geq
\frac{\alpha}{2(1-\alpha)}\ln\frac{\kappa+1}{d+1}.
$$

Note that the function $f(x)=-\ln x$ is convex.
Concerning the min-entropy $R_{\infty}(\textbf{p})=-\ln(\max p_{j})$, we have
$$
\frac{1}{d+1}\sum_{b=1}^{d+1}R_{\infty}(\mathcal{P}^{(b)}|\rho)\geq-\ln\left[\frac{1}{d+1}\sum_{b=1}^{d+1}(\max_{1\leq n\leq d}p_{n}^{(b)})\right],
$$
where $p_{n}^{(b)}=\mathrm{Tr}(P_{n}^{(b)}\rho)$ and $\mathcal{P}^{(b)}=\{P_{n}^{(b)}\}_{n=1}^{d}, b=1,2,\ldots,d+1$.
Define the function
$$
g_{d}(x)=d^{-1}(1+\sqrt{d-1}\sqrt{xd-1}),
$$
which is concave and increasing. It has been proved that \cite{RastEur},
$$
\frac{1}{d+1}\sum_{b=1}^{d+1}(\max_{1\leq n\leq d}p_{n}^{(b)})\leq\frac{1}{d+1}\sum_{b=1}^{d+1}g_{d}\left(C^{(b)}(\kappa,\rho)\right),
$$
where $C^{(b)}(\kappa,\rho)=\sum_{n=1}^{d}(p_{n}^{(b)})^{2}$. Using the concavity of $g_{d}(x)$, we have
$$
\frac{1}{d+1}\sum_{b=1}^{d+1}g_{d}\left(C^{(b)}(\kappa,\rho)\right)\leq g_{d}\left(\frac{C(\kappa,\rho)}{d+1}\right),
$$
where $C(\kappa,\rho)$ is the index of coincidence of the set of MUMs.
Thus we obtain a state-dependent uncertainty relation for MUMs in terms of min-entropy:
$$
\frac{1}{d+1}\sum_{b=1}^{d+1}R_{\infty}(\mathcal{P}^{(b)}|\rho)\geq-\ln g_{d}\left(\frac{C(\kappa,\rho)}{d+1}\right).
$$
Note that the function $-\ln g_{d}(x)$ is decreasing and $\mathrm{Tr}(\rho^{2})\leq1$, we have the following state-independent uncertainty relation
$$
\frac{1}{d+1}\sum_{b=1}^{d+1}R_{\infty}(\mathcal{P}^{(b)}|\rho)\geq-\ln g_{d}\left(\frac{\kappa+1}{d+1}\right).
$$

For $\alpha>0$ and $\alpha\neq1$, the Tsallis entropy of probability distribution $\textbf{p}=(p_{1},p_{2},\ldots p_{d})$ is defined as \cite{Tsa}
$$
H_{\alpha}(\textbf{p})=\frac{1}{1-\alpha}(\sum_{j=1}^{d}p_{j}^{\alpha}-1).
$$
Define the $\alpha$-logarithm for $x>0$ as
$$
\ln_{\alpha}(x)=\frac{x^{1-\alpha}-1}{1-\alpha},
$$
the Tsallis entropy can be rewritten as
$$
H_{\alpha}(\textbf{p})=-\sum_{j=1}^{d}p_{j}^{\alpha}\ln_{\alpha}(p_{j})=\sum_{j=1}^{d}p_{j}\ln_{\alpha}(\frac{1}{p_{j}}).
$$
When $\alpha\rightarrow1$, the $\alpha$-logarithm is reduced to $\ln x$, and $H_{1}(\textbf{p})$ is just the Shannon entropy.

For $\alpha\in(0,2]$, it has been proved that \cite{RastEur}
$$
H_{\alpha}(\textbf{p})\geq\ln_{\alpha}(\frac{1}{C(\textbf{p})}),
$$
where $C(\textbf{p})$ is the index of coincidence of $\textbf{p}$. Using the convexity of the function $f(x)=\ln_{\alpha}(\frac{1}{x})$ \cite{RastEur}, we have the following state-dependent uncertainty relation for MUMs in terms of Tsallis entropy
$$
\frac{1}{d+1}\sum_{b=1}^{d+1}H_{\alpha}(\mathcal{P}^{(b)}|\rho)\geq
\ln_{\alpha}(\frac{d+1}{C(\kappa,\rho)}),
$$
where $0<\alpha\leq2.$ The state-independent form is given by
$$
\frac{1}{d+1}\sum_{b=1}^{d+1}H_{\alpha}(\mathcal{P}^{(b)}|\rho)\geq
\ln_{\alpha}(\frac{d+1}{\kappa+1}),
$$
since the function $f(x)=\ln_{\alpha}(\frac{1}{x})$ is decreasing for $0<\alpha\leq2$ and $\mathrm{Tr}(\rho^{2})\leq1$.

\section{Conclusion}
We have formulated uncertainty relations related to the mutually unbiased measurements.
We have presented a number of inequalities and derived the lower entropic bounds by calculating the index of coincidence for MUMs.
Both state-dependent and state-independent inequality forms have been given. Furthermore, we have considered the
uncertainty relations for MUMs in terms of R\'{e}nyi and Tsallis entropies.
These entropies have been wildly used in quantum information theory, especially in studying uncertainty relations \cite{RastEur}.
The results presented in this work depend on the parameter $\kappa$ of MUMs. The lower entropic bounds become tighter when $\kappa$ increases.

\vspace{2.5ex}
\noindent{\bf Acknowledgments}\, \,
This work is supported by the NSFC under number 11275131.

\end{document}